\documentclass[twocolumn,showpacs,preprintnumbers,
superscriptaddress,amsmath]{revtex4}

\newcommand{\NPA}[3]{Nucl.\ Phys.\ {\bf A#1},\ #2 (#3)}
\newcommand{\NPB}[3]{Nucl.\ Phys.\ {\bf B#1},\ #2 (#3)}

\newcommand{\PLB}[3]{Phys.\ Lett.\ B\ {\bf #1},\ #2 (#3)}

\newcommand{\PRL}[3]{Phys.\ Rev.\ Lett.\ {\bf #1},\ #2 (#3)}

\newcommand{\PRD}[3]{Phys.\ Rev.\ D\ {\bf #1},\ #2 (#3)}

\renewcommand\a{\alpha}
\renewcommand\b{\beta}
\newcommand\g{\gamma}

\newcommand\m{\mu}

\newcommand\p{\pi}

\newcommand{\be}{\begin{equation}}
\newcommand{\ee}{\end{equation}}
\newcommand{\bea}{\begin{eqnarray}}
\newcommand{\eea}{\end{eqnarray}}
\newcommand{\ba}[1]{\begin{array}{#1}}
\newcommand{\ea}{\end{array}}

\begin{document}

\title{Electromagnetic Meissner effect in spin-one color superconductors}

\author{Andreas Schmitt}
\email{aschmitt@th.physik.uni-frankfurt.de}
\affiliation{Institut f\"ur Theoretische Physik, 
J.W. Goethe-Universit\"at, D-60054 Frankfurt/Main, Germany}

\author{Qun Wang}
\email{qwang@th.physik.uni-frankfurt.de}
\affiliation{Institut f\"ur Theoretische Physik, 
J.W. Goethe-Universit\"at, D-60054 Frankfurt/Main, Germany}
\affiliation{Physics Department, Shandong University, Jinan, 
Shandong, 250100, P.R. China}

\author{Dirk H. Rischke}
\email{drischke@th.physik.uni-frankfurt.de}
\affiliation{Institut f\"ur Theoretische Physik, 
J.W. Goethe-Universit\"at, D-60054 Frankfurt/Main, Germany}

\date{\today}

\begin{abstract}
It is shown that color-superconducting quark matter,
where quarks of the same flavor form Cooper pairs with spin one,
exhibits an electromagnetic Meissner effect. 
This is in contrast to spin-zero color superconductors where
Cooper pairs consist of quarks with different flavors.
\end{abstract}

\maketitle 

Quantum chromodynamics (QCD) is an asymptotically free theory
\cite{asymp} and, thus, quark matter
at large quark chemical potential
$\mu$ is a weakly coupled system. In this case,
the dominant interaction between two quarks
is single-gluon exchange. Single-gluon exchange is attractive in the 
color-antitriplet channel. Consequently, at sufficiently low
temperatures, the quark Fermi surface is unstable with respect
to the formation of Cooper pairs. Since this is analogous to what happens
in ordinary superconductors \cite{bcs}, this phenomenon was
termed {\em color\/} superconductivity \cite{bailin}.

Color superconductivity was studied 
from first principles in the framework of QCD at weak coupling
\cite{QCDgapeq,rischke1} as well as in more phenomenological 
Nambu--Jona-Lasinio-type models \cite{alford1}. Both approaches indicate
that the color-superconducting state is the true ground state of
quark matter at any density beyond the quark-hadron phase transition
and at sufficiently low temperature.
They also agree in the magnitude of the color-superconducting
gap parameter, $\phi$, which they predict to be of the order of tens of MeV
(for quark Cooper pairs with total spin zero) at densities of the
order of ten times the nuclear matter ground state density.
Gap parameters of this order of magnitude may have 
enormous phenomenological implications, since
the transition temperature to the normal conducting phase
is typically of the order of $\phi$ \cite{rischke1,wang,schmitt}. 
For instance, during the evolution of a neutron star,
the temperature ranges from a few tens of MeV down to a few keV
\cite{pons}.
If its core is sufficiently dense to consist of quark matter, 
this quark matter core is then very likely a color superconductor.

Color superconductivity is more complex than ordinary
superconductivity, because quarks do not only carry 
electric, but also color and flavor charge.
Two different quark flavors may form Cooper pairs in the
color-antitriplet, flavor-singlet, total spin-zero channel 
(the so-called 2SC phase) \cite{bailin}.
For three different quark flavors, the favored state is 
the so-called color-flavor locked (CFL) phase \cite{alford2}, with
spin-zero Cooper pairs in the color-antitriplet, flavor-antitriplet 
representation.

A necessary condition for pairing of quarks of different flavor 
is that their respective Fermi surfaces are equal. In physical systems,
however, this condition may be hard to achieve. For instance, 
compact stellar objects are neutral with respect to electric and color
charge. This requires the introduction of separate chemical potentials
for quarks which differ in color and flavor \cite{neutrality}. 
Imposing the constraints of color and electric charge neutrality 
leads to different values for these chemical potentials.
This effect and the mass difference between
the light up and down quarks and the heavier strange quark 
\cite{berges} may then lead to different Fermi surfaces for each quark 
species. If this difference is of the order of, or larger than, the
color-superconducting gap parameter $\phi$, a color-superconducting
state where quarks with different flavor form spin-zero pairs with zero
momentum is no longer favored. Then, besides a transition to the normal
conducting state \cite{bedaque}, there are at least four other possibilities.
The first two possibilities are either a displacement \cite{bowers} or
a deformation \cite{muether} of the Fermi spheres of the two
quark species forming the Cooper pair, breaking translational
or rotational invariance, respectively.
The third possibility is an interior gap structure for
the quark species with the larger Fermi momentum \cite{liu}.

The fourth possibility is that quarks of the same flavor form
Cooper pairs with total spin one 
\cite{bailin,iwasaki,rischke1,schaefer,alford3}. 
Quarks with the same flavor have the same mass and the same electric charge,
and thus pairing is neither affected by a mass difference nor a
nonzero electric chemical potential which may be required to fulfill 
the constraint of electric neutrality. Moreover,
a potentially nonzero color chemical potential does not destroy
the Cooper pairs, because we expect it to be parametrically much
smaller than the gap, $\mu_{\rm color} \sim \phi^2/(g \mu) \ll \phi$, 
where $g$ is the strong coupling constant \cite{colorneutrality}. 
(A color chemical potential is
necessary in models which treat color as a global charge.
In QCD, the role of the color chemical potential is assumed
by a constant gluon background field $A^0_a$
\cite{colorneutrality}.)

According to the results of Refs.\ \cite{schmitt,schaefer},
the gap in spin-one color superconductors is of the order of 
20 - 400 keV (assuming that the gap in the 2SC phase is of the order of
10 - 100 MeV). The critical temperature 
is therefore of the order of 10 - 400 keV \cite{schmitt}.
Consequently, when the core of a neutron star cools
below this temperature, it could very well 
consist of quark matter in a spin-one
color-superconducting state.

The interesting question is how such a state affects the properties
of the star and whether this leads to observable consequences.
The best known properties of a neutron star are its radius and mass,
which are determined by the equation of state. 
A recent study \cite{alfordreddy} shows that radii and masses of 
compact stellar objects with a color-superconducting quark matter core
do not change appreciably from the values expected for ordinary
neutron stars.
Another property of a neutron star with potentially observable
consequences is its magnetic field.
Due to an admixture of protons in neutron matter, 
the core of an ordinary neutron star is a superconductor, and
magnetic fields experience the Meissner effect.
However, if the core of a neutron star is a spin-zero color
superconductor (for instance in the 2SC or CFL phase), 
there is no electromagnetic Meissner effect \cite{emprops}. 
Since charge neutrality may favor a spin-one 
over a spin-zero color-superconducting state,
a natural question is whether the electromagnetic Meissner 
effect is also absent in a spin-one color superconductor.
In order to answer this question, in this letter
we study the pattern of symmetry breaking of the local gauge symmetries, 
and then also present
the results of an explicit calculation of the Meissner masses.

In ordinary superconductors, the electromagnetic gauge group
$U(1)_{em}$ is broken due to the fact that the electrons in a Cooper
pair carry electric charge. This leads to the electromagnetic
Meissner effect, i.e., magnetic fields only penetrate a finite
distance into the superconductor. The inverse distance can
be associated with a nonzero photon mass, the so-called Meissner mass.
Since quarks are not only electrically charged but also carry color,
besides the electromagnetic  $U(1)_{em}$ symmetry
also the $SU(3)_c$ gauge symmetry of the strong interaction
is broken in a color superconductor. This leads to the
color Meissner effect, i.e., the gluons obtain Meissner masses and
color-magnetic fields are expelled.
The question is whether all eight gluons and the photon become
massive. This depends on the particular pattern of how the local
symmetries are broken in the color superconductor. If there is a residual
local symmetry, the corresponding gauge bosons remain massless.
This residual symmetry group of $SU(3)_c\times U(1)_{em}$ leaves 
the gap matrix $\Delta$ invariant,
\be \label{invariant}
(g_c\times g_{em}) \Delta (g_c^T\times g_{em}^T) \stackrel{!}{=} \Delta \,\, ,
\ee   
where $g_c\in SU(3)_c$, $g_{em}\in U(1)_{em}$, and $T$ denotes the
transpose.
In general, the gap matrix $\Delta$ is a matrix in 
color, flavor, and Dirac (spin) space
\cite{bailin,QCDgapeq,rischke1,alford1}.
Since pairing occurs in the attractive color-antitriplet channel,
the color structure of the gap matrix 
corresponds to the color-antitriplet
${\bf\bar{3}}_c$ representation of the $SU(3)_c$ gauge group. 
In a spin-one color superconductor, 
the spin structure of the gap matrix corresponds to the symmetric spin
triplet ${\bf 3}_J$ representation of the $SU(2)_J$ spin group,
which is also a representation of $SO(3)_J$.
The gap matrix is diagonal in flavor space,
since quarks in a Cooper pair carry the same flavor.
For the moment, let us consider quark matter with a 
single flavor only, $N_f=1$.
The case of several quark flavors (where
each flavor pairs at its respective Fermi surface)
will be discussed below. The gap matrix can be written as
\be
\Delta=\Phi_a^iJ_a\otimes v^i \,\, ,
\ee
where $J_a$ and $v_i$ $(a,i=1,2,3)$ are bases of ${\bf\bar{3}}_c$ 
and ${\bf 3}_J$, respectively, and $\Phi_a^i$ is the order parameter.
The form of the order parameter defines the phase of the condensate.
As in helium-3, there is a multitude of possible phases
for spin-one condensates \cite{vollhardt}. In this letter we 
only consider the polar phase and the color-spin locked (CSL) phase 
\cite{schmitt,schaefer}. The order parameters in these phases are
\be
\left(\Phi_a^i\right)_{polar}\sim\delta_{a3}\delta^{i3} \,\, ,
\;\;\;\;
\left(\Phi_a^i\right)_{CSL}\sim\delta_a^i \,\, .
\ee

In the polar phase, the condensate points in a 
fixed direction in real space, 
which breaks the global spatial symmetry group $SO(3)_J$ 
to $SO(2)_J$. The condensate also points in a fixed direction in
color space, which spontaneously breaks the local $SU(3)_c$ 
gauge symmetry to a residual $SU(2)_c$ gauge group. From 
Eq.\ (\ref{invariant}) we
deduce that the residual subgroup which leaves the order parameter
invariant is generated by the three generators of $SU(2)_c$
(corresponding to $T_1,\, T_2,\, T_3$ of the original $SU(3)_c$)
and the generator $\tilde{Q}_{polar}=Q-2\sqrt{3}q\,T_8$,
where $T_8$ is one of the generators of $SU(3)_c$ and $Q\equiv q{\bf 1}_J$
is the generator of $U(1)_{em}$. 
The constant $q$ is the electric charge 
of the single quark flavor considered here
(2/3 for $u$ quarks and $-1/3$ for $d$ or $s$ quarks).
The generator $Q$ is proportional to the unit matrix
in spin triplet space, ${\bf 1}_J$, since all
states of the spin triplet have the same electric charge.
Consequently, the symmetry breaking pattern is
$SU(3)_c\times U(1)_{em} \to SU(2)_c \times \tilde{U}(1)$,
where $\tilde{U}(1)$ is generated by $\tilde{Q}_{polar}$. The
existence of a nontrivial residual gauge symmetry is equivalent to the fact 
that there are charges with respect to which the Cooper pairs are
neutral. These are the two color charges corresponding to the
$SU(2)_c$ gauge symmetry and the $\tilde{Q}_{polar}$ charge
corresponding to the $\tilde{U}(1)$ gauge symmetry.
The gauge boson of the latter is a superposition of the
photon and the eighth gluon of $SU(3)_c$.
This superposition is mathematically given by an orthogonal
rotation of the original gauge fields by an angle $\theta$.
In general, a generator $\tilde{Q} = Q + \eta T_8$
results in a mixing angle given by
$\cos^2\theta = g^2/(g^2 + \eta^2 e^2)$, 
where $e$ is the electromagnetic coupling constant \cite{emprops}. 
In our case the mixing angle, $\theta_{polar}$, is determined by
this expression with $\eta=-2\sqrt{3}\,q$.
Since $e \ll g$, the mixing angle is small,
$\theta_{polar} \simeq 2 \sqrt{3} q e/g \sim q/3$. Consequently, 
the main contribution to the gauge boson of the local $\tilde{U}(1)$ 
symmetry comes from the original photon, with a small admixture of the 
eighth gluon. 
This justifies to call this gauge boson the ``new'' photon.
There is no electromagnetic Meissner effect, since the 
new photon can penetrate the color-superconducting phase.
This is similar to other color-superconducting phases, for
instance the 2SC phase or 
the CFL phase \cite{emprops}. In both cases, there
is a $\tilde{U}(1)$ gauge symmetry and thus no electromagnetic Meissner
effect.

In the CSL phase, the order parameter
breaks $SU(3)_c\times SO(3)_J$ to the diagonal subgroup
$SO(3)_{c+J}$ \cite{schaefer}. This is analogous
to the breaking of color-flavor symmetries in the CFL phase,
where $SU(3)_c \times SU(3)_f \rightarrow SU(3)_{c+f}$.
The residual $SO(3)_{c+J}$ and $SU(3)_{c+f}$ groups are global
symmetries, and thus not gauged.
This similarity between CSL and CFL phases does, however, not extend
to the behavior concerning electromagnetism.
Unlike the CFL phase, it turns out that for the CSL phase there is 
{\em no\/} nontrivial subgroup of 
$SU(3)_c\times U(1)_{em}$ that leaves the gap matrix invariant, i.e.,
the only possible solution to Eq.\ (\ref{invariant}) is $g_c=g_{em}={\bf 1}$.
Consequently, the symmetry breaking pattern is
$SU(3)_c\times U(1)_{em} \to \bf{1}$.
This is equivalent to the fact that a Cooper pair 
in the CSL phase is {\em not\/} neutral, neither with respect
to ordinary electric charge nor with respect to any possible 
new combination of color and electric charge. This fact has the
physical consequence that there is an electromagnetic
Meissner effect for the CSL phase \cite{schaefer}. 
  
In the following, we confirm the above qualitative arguments by
an explicit calculation of the Meissner masses. 
The Meissner mass is defined as the zero-energy, zero-momentum
limit of the spatial $(ii)$ components of the polarization tensor
\cite{meissner}
\be \label{defmeissner}
m_{ab}^2\equiv \mbox{lim}_{p\to 0} \Pi_{ab}^{ii}(0,p) \,\, ,
\;\;\;\; a, b=  1, \ldots, 8, 9 \,\, ,
\ee 
where the first eight indices correspond to the gluons, and the
ninth index to the photon, $9 \equiv \g$.
If there is mixing of gluons and photons,
the $9 \times 9$ Meissner mass matrix $m^2_{ab}$ is not diagonal.
The Meissner masses for the physically relevant modes
are obtained by diagonalizing this matrix.
In weak coupling, the polarization tensor may be computed in
one-loop approximation. At zero temperature, only the quark
loop contributes. Our calculation follows
the method of Ref.\ \cite{meissner};
details are deferred to a future paper \cite{future}. 
It turns out that the gluon part of the mass matrix is
diagonal, $m_{ab}^2 \equiv \delta_{ab} m_{aa}^2$ for $a,b = 1, \ldots, 8$.
In Table \ref{table1} we collect all results. 

\begin{table}
\begin{tabular}{|c||cccccccc|cc|c|}
\hline
& \multicolumn{8}{c}{$m^2_{aa}$}\vline & 
\multicolumn{2}{c}{$m^2_{a\g}$}\vline
&  $m_{\g\g}^2$ \\
\hline\hline
$a$ & 1 & 2 & 3 & 4 & 5 & 6 & 7 & 8 & 1-7 & 8 & 9 \\ 
\hline
polar & 0 & 0 & 0 & $\frac{1}{2}$ & $\frac{1}{2}$ & $\frac{1}{2}$ & 
$\frac{1}{2}$ & $\frac{1}{3}$ & 0 &   $\frac{2}{\sqrt{3}}q$ &  $4q^2$ \\
\hline   
CSL & $\b$ & $\a$ & $\b$ & $\b$ & $\a$ & $\b$ & $\a$ & $\b$ & 0 &  0 
& $6 q^2$ \\
\hline 
\end{tabular} 
\caption{Zero-temperature masses (squared) calculated 
for the polar phase and the CSL phase of a spin-one color
superconductor. The gluon masses are
given in units of $g^2\mu^2/(6\pi^2)$, the mixed masses
in units of $eg \m^2/(6 \p^2)$, and the photon masses
in units of $e^2 \m^2/(6 \p^2)$. The constants $\a$ and $\b$
are defined as $\a \equiv (3+4\ln 2)/27$, $\b \equiv (6-4\ln 2)/9$.
}\label{table1}
\end{table}
  
For the polar phase, the vanishing masses for gluons 1, 2, and 3 
indicate the unbroken $SU(2)_c$ subgroup.
The nonzero Meissner mass $m_{8\g}^2$ reflects the mixing of 
eighth gluon field, $A_8$, and the photon, $A_9 \equiv A_\gamma$.
In terms of the physically relevant modes, 
$\tilde{A}_a$, the Meissner mass matrix is diagonal,
\be
\sum_{ab} A_a m_{ab}^2 A_b \equiv \sum_a 
\tilde{A}_a \tilde{m}_{a}^2 \tilde{A}_a\,\,.
\ee
For $a=1,\ldots,7$, $\tilde{A}_a \equiv A_a$ (and,
correspondingly, $\tilde{m}_a^2 \equiv m_{aa}^2$), whereas
the new gluon, $\tilde{A}_8$, and the new photon, $\tilde{A}_\g$,
are obtained by an orthogonal rotation of $A_8$ and $A_\gamma$,
\be
\left( \begin{array}{c} \tilde{A}_8 \\ \tilde{A}_\g \end{array}
\right) = \left( \begin{array}{cc} \cos \vartheta & \sin \vartheta \\
                                   -\sin \vartheta & \cos \vartheta 
                  \end{array} \right)
\left( \begin{array}{c} A_8 \\ A_\g \end{array} \right)\,\, ,
\ee 
where $\tan (2\vartheta) \equiv 2\, m_{8\g}^2/(m_{88}^2-m_{\g\g}^2)$.
With the numbers of Table \ref{table1} we find that the
rotation angle $\vartheta$ is identical to the mixing angle
$\theta_{polar}$ found in the above group-theoretical argument. 
The Meissner masses for the rotated gauge fields are
\be
\tilde{m}^2_{8}= \left(\frac{1}{3}g^2+4q^2e^2 \right) \,
\frac{\m^2}{6\pi^2}\,\, , \qquad
\tilde{m}^2_{\g}=0 \,\, .
\ee
The massless new photon confirms that there is no electromagnetic 
Meissner effect in the
polar phase, whereas there is a color Meissner effect for the new gluon.

For the CSL phase, the particular pattern for the gluon masses
reflects the residual global $SO(3)_{c+J}$ symmetry which is generated by a 
combination of the antisymmetric Gell-Mann matrices $T_2$, $T_5$, and $T_7$ 
and the generators of $SO(3)_J$.
In contrast to the polar phase, now all mixed masses are zero,
indicating that there is no mixing between gluons and the photon.
All gauge bosons have a nonzero  Meissner mass. This coincides with 
the above group-theoretical argument.

Let us now discuss the situation with more than one quark flavor,
where the quarks of each flavor separately form spin-one Cooper pairs.
In general, each quark flavor has a different
chemical potential, $\mu_i$, $i = 1, \ldots, N_f$.
In this case, the results of Table \ref{table1} have
to be modified. All gluon Meissner masses $m_{aa}^2$ 
have to be  multiplied by a factor $\sum_i (\mu_i/\mu)^2$.
In the mixed masses $m_{a\g}^2$ 
we have to replace the single quark electric charge $q$ by a
factor $\sum_i q_i (\mu_i/\mu)^2$. Finally, one has to substitute
the square of the charge $q^2$ in the photon masses $m _{\g\g}^2$ 
by a factor $\sum_i (q_i \mu_i/\mu)^2$.

For the polar phase, these modifications have the following
effect. Only if $\sum_{i,j} q_i (q_i - q_j) (\mu_i \mu_j)^2 =
0$, there is a massless combination of photon and gluon. For a system with two
quark flavors, for instance $u$ and $d$, or $u$ and $s$, this condition
is equivalent to $\mu_u^2 \mu_d^2 = 0$, or $\mu_u^2 \mu_s^2 =0$,
respectively. This means that for these two combinations of flavors,
there is always a Meissner effect, unless the chemical potential
of one of the two quark flavors is zero, i.e., this
flavor is absent. For a two-flavor system consisting of $d$ and
$s$ quarks, however, the above condition is always trivially
fulfilled, i.e., there is no Meissner effect in this case. This is
to be expected, since $d$ and $s$ quarks carry the same electric
charge and thus appear as one single flavor with respect to
electromagnetic interactions.
For three flavors, the above condition is equivalent to
$\mu_u^2 (\mu_d^2 + \mu_s^2) = 0$, and can only be fulfilled if
either $\mu_u=0$, or $\mu_d = \mu_s = 0$. Both cases have been
discussed above. 
We therefore conclude that, for more than
one quark flavor, there is an electromagnetic Meissner effect
in the polar phase, unless all quarks carry the same electric charge.

In the CSL phase, the mixed masses always vanish, while the results
for the gluon masses and the photon mass
are modified by the same factors as for the polar phase. 
Thus, the conclusion that there is no
photon-gluon mixing and an electromagnetic Meissner effect in the
CSL phase remains valid also for more than one quark flavor.

While spin-zero color superconductors could be of type II
at small $\mu$ \cite{blaschke,baym},
a spin-one color superconductor is most likely always of type I,
because the ratio of the penetration depth to the coherence length
is of order $\sim \phi/(e \mu) \sim 10^{-3}\, [100\, {\rm MeV}/(e\mu)] \ll 1$.
Consequently, the magnetic field is completely expelled from the 
core of a compact stellar object, if it is
a spin-one color-superconductor.
This is true unless the magnetic field exceeds the critical field
strength
for the transition to the normal conducting state. The magnetic
field in neutron stars is typically of the order of $10^{12}$ Gauss.
This is much smaller than the critical magnetic field which,
from the results of Ref.\ \cite{baym}, we estimate to be of
the order of $10^{16}$ Gauss. 

In conclusion, a compact stellar object with a core consisting of
quark matter in the spin-one color-superconducting state 
is, with respect to its electromagnetic properties, 
different from an ordinary neutron star: ordinary neutron star matter is 
commonly believed to be an electromagnetic superconductor of type II,
while a spin-one color superconductor is of type I. 
The question whether neutron star
matter is a type-I or type-II superconductor has recently
stirred a lot of attention \cite{zhit}, because it was shown that type-II
superconducting matter is incompatible with
the observation of pulsars with precession periods
of order 1 year \cite{link}. The presence of
spin-one color-superconducting quark matter in the pulsars' core
could explain this observation.

{\bf Acknowledgment.} We would like to thank M.\ Alford,
M.\ Lang, K.\ Rajagopal, T.\ Sch\"afer, and I.A.\ Shovkovy 
for comments and discussions.
Q.W.\ acknowledges financial support from the
Alexander von Humboldt-Foundation.

\end{document}